\documentclass[osajnl,twocolumn,showpacs,superscriptaddress,10pt]{revtex4-1} 
\usepackage{amsmath,amssymb,graphicx}

\usepackage{mathtools}
\usepackage{xfrac}

\begin{document}

\title{Competition between Raman and Kerr effects in microresonator comb generation}

\author{Yoshitomo Okawachi}\email{Corresponding author: y.okawachi@columbia.edu}
\affiliation{Department of Applied Physics and Applied Mathematics, Columbia University, New York, NY 10027}

\author{Mengjie Yu}
\affiliation{School of Applied and Engineering Physics, Cornell University, Ithaca, NY 14853}
\affiliation{Department of Applied Physics and Applied Mathematics, Columbia University, New York, NY 10027}

\author{Vivek Venkataraman}
\affiliation{School of Engineering and Applied Sciences, Harvard University, Cambridge, MA 02138}

\author{Pawel M. Latawiec}
\affiliation{School of Engineering and Applied Sciences, Harvard University, Cambridge, MA 02138}

\author{Austin G. Griffith}
\affiliation{School of Applied and Engineering Physics, Cornell University, Ithaca, NY 14853}

\author{Michal Lipson}
\affiliation{Department of Electrical Engineering, Columbia University, New York, NY 10027}

\author{Marko Lon\v{c}ar}
\affiliation{School of Engineering and Applied Sciences, Harvard University, Cambridge, MA 02138}

\author{Alexander L. Gaeta}
\affiliation{Department of Applied Physics and Applied Mathematics, Columbia University, New York, NY 10027}

\begin{abstract}We investigate the effects of Raman and Kerr gain in crystalline microresonators and determine the conditions required to generate modelocked frequency combs. We show theoretically that strong, narrowband Raman gain determines a maximum microresonator size allowable to achieve comb formation. We verify this condition experimentally in diamond and silicon microresonators and show that there exists a competition between Raman and Kerr effects that leads to the existence of two different comb states.
\end{abstract}


\maketitle 


Over the past decade, there has been significant development of microresonator-based frequency combs based on four-wave mixing (FWM) parametric oscillation \cite{Kippenberg,Del'Haye,Savchenkov,Levy,Razzari,Ferdous,Papp,Jung,Hausmann,Xue,Huang,Yu,Yi,Karpov}, with applications including spectroscopy, metrology, and wavelength division multiplexing \cite{Diddams07,Udem,LevyPTL,Koos}. Operating in the anomalous group-velocity dispersion (GVD) regime of the microresonator allows for modelocking and temporal cavity soliton formation \cite{Saha,Herr,Yi15,Joshi}, which enables applications requiring frequency precision. Studies have also explored the effects of Raman scattering on parametric oscillation and comb generation \cite{Min,Agha,Grudinin,Liang,Griffith,Latawiec,Kumar,Liu,Yang}. For example, self-frequency shifts have been observed in amorphous materials such as silica and silicon nitride \cite{Yi,Karpov}, where the presence of a red-shifted Kerr soliton has been attributed to the Raman effect. Furthermore, Raman lasing and Raman-FWM interactions have been observed in crystalline structures such as magnesium fluoride, silicon, diamond, and aluminum nitride \cite{Grudinin,Liang,Griffith,Latawiec,Liu}, resulting in non-trivial nonlinear interactions in the microresonator and the disruption of soliton formation.


\begin{figure}[t]
\centering
\centerline{\includegraphics[width=6.0cm]{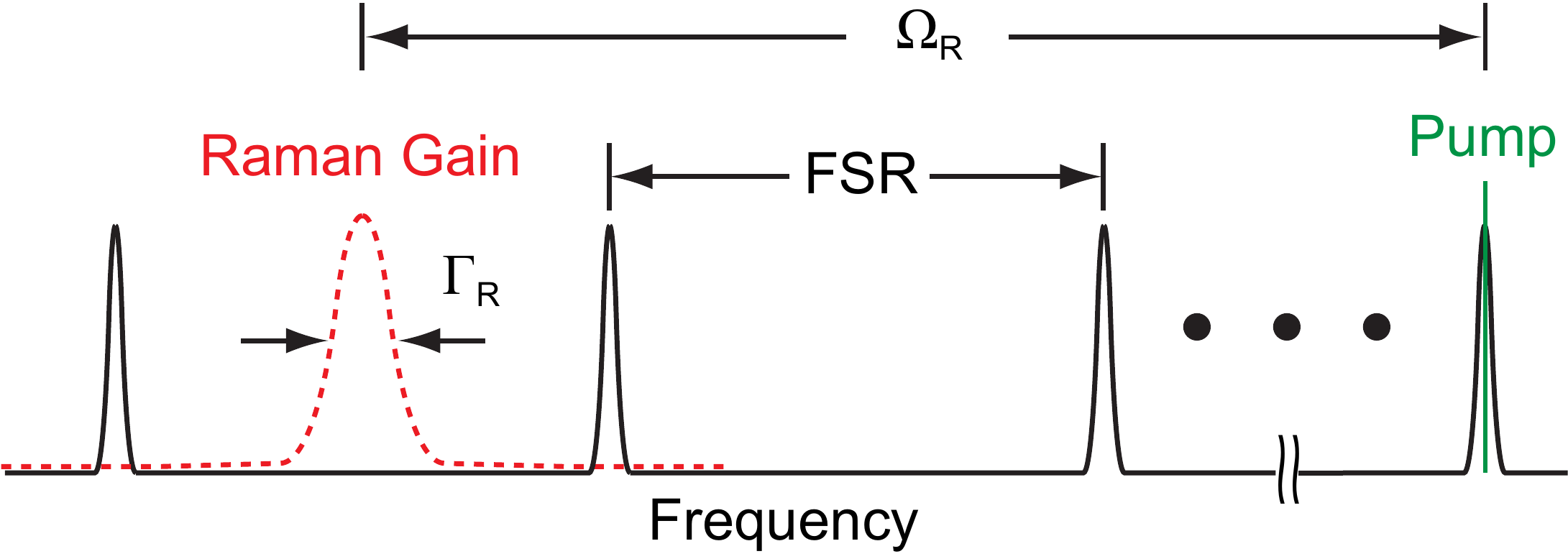}}
\caption{Scheme for suppression of effective Raman gain in microresonators. The free spectral range is chosen such that when the peak of the Raman gain lies between two adjacent cavity resonances, the nearest Stokes cavity mode is sufficiently far detuned from the gain peak. $\Omega_R$ is the Raman frequency shift (\emph{eg}. 40 THz for diamond and 15.6 THz for silicon).}
\label{Fig1}
\end{figure}

\begin{figure*}[t]
\centerline{\includegraphics[width=15.0cm]{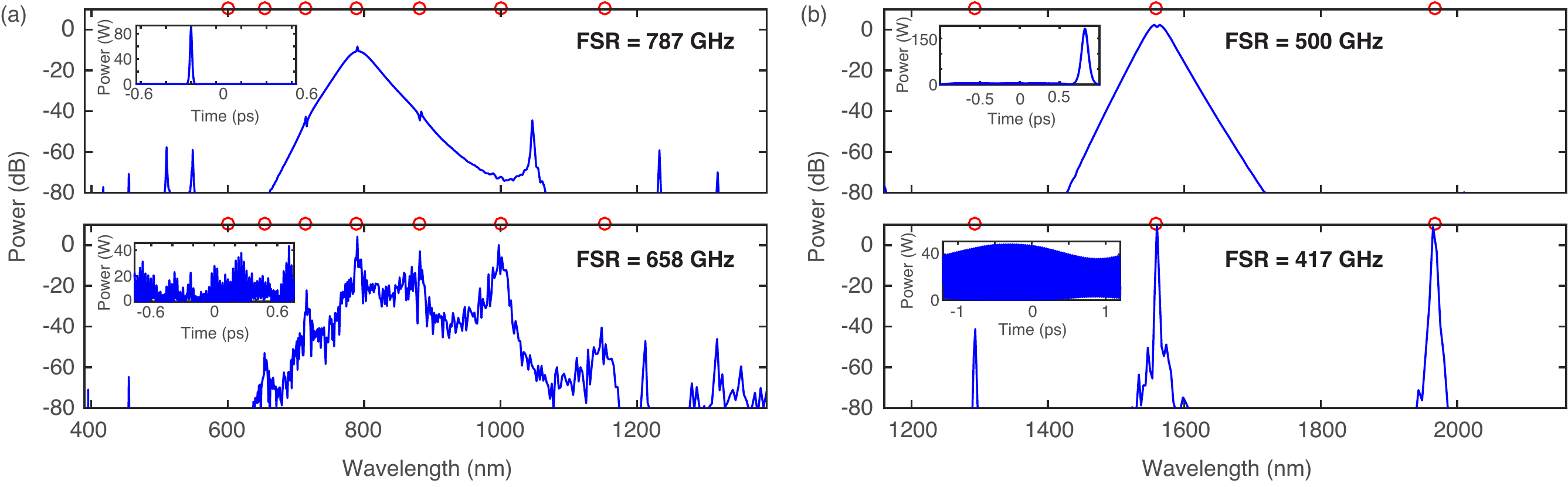}}
\caption{(a) Simulated spectrum for comb generation in diamond microresonators pumping at 790 nm for FSR's of 787 GHz (top) and 658 GHz (bottom). (b) Simulated spectrum a 1560 nm pump for FSR's of 500 GHz (top) and 417 GHz (bottom). The insets show the corresponding temporal profile. The red circles indicate the Raman frequency shift.}
\label{Fig2}
\end{figure*}

In this paper, we theoretically investigate the combined effects of Raman and parametric gain on Kerr comb generation in diamond and silicon microresonators. Diamond is an ideal platform for Kerr combs since it has a large Kerr nonlinearity ($n_2 = 1.3\times10^{-19}$ m$^2$/W), enables broadband parametric gain with dispersion engineering of the waveguides, and has a wide optical transparency window covering the UV to the mid-infrared (mid-IR) \cite{Hausmann}. It also has a strong Raman gain ($g_R = 26$ cm/GW at 800 nm) with a narrow linewidth ($\Gamma_R/2\pi = 60$ GHz). Silicon is a well established photonics platform and offers promise for frequency comb applications in the mid-IR  \cite{Yu, Griffith, Yu_arXiv}. It has an even larger nonlinearity than diamond ($n_2 = 3\times10^{-18}$ m$^2$/W at 3100 nm) \cite{Gai} and we estimate a Raman gain $g_R = 3$ cm/GW for a Stokes field at 3700 nm \cite{Raghunathan,Pask}, with a linewidth $\Gamma_R/2\pi = 105$ GHz. Since the Raman gain is large and inherent in these crystalline materials, it can play an integral part in the nonlinear dynamics in the microresonator. We show that it is possible to minimize the strength of the Raman effect by designing the device geometry. Our analytic and numerical studies show that tuning the free-spectral range (FSR) of the microresonator can be used to achieve a Kerr comb state or a Raman oscillation state. For Kerr comb generation, the FSR must be chosen such that the peak of the Raman gain is far detuned from any cavity resonance in order to suppress the Raman gain as compared to the parametric gain. This condition posts a maximum limit of the microresonator's dimension (or round-trip path length), especially for diamond microresonators operating in the visible regime where the Raman gain is typically much larger. In addition, we investigate the competing effects near the FSR limit in silicon microresonators. Our analysis offers a novel route to achieving Kerr comb generation in the presence of strong Raman effects and can be generalized to other nonlinear crystalline materials with a narrow Raman gain feature. 

First, we analyze competing Raman and Kerr effects in crystalline microresonators. The parametric gain can be expressed as $G_{NL} = g_KPL/A$, where the parametric gain coefficient is $g_K = 4\pi n_2/\lambda$ where $\lambda$ is the pump wavelength, $n_2$ is the nonlinear index coefficient, $P$ is the pump power, and $A$ is the effective optical mode area. Similarly, the Raman gain can be written as $G_R = g_RPL/A$. In order to achieve parametric oscillation and Kerr comb formation in a crystalline microresonator, the Raman gain must then be suppressed by more than the ratio$g_R/g_K$ between the two gain coefficients. By assuming that the Raman spectral gain profile can be expressed as a Lorentizan $g_R(\delta) = g_R/(1+4\delta^2/\Gamma_R^2)$ and choosing the FSR such that the peak Raman gain is centered between two cavity modes (see Fig. \ref{Fig1}), the effective gain at the nearest Stokes cavity mode will be suppressed by more than $g_R/g_K$ if $\delta/\Gamma_R > \frac{\sqrt{g_R/g_K}}{2}$. Thus positioning the peak Raman gain between two cavity resonances (Fig. 1) is critical for pure Kerr comb generation. 

In diamond, for a pump wavelength of 800 nm, the Kerr and Raman gain coefficients are $g_K = 0.2$ cm/GW and $g_R = 26$ cm/GW \cite{Savitski}. To achieve pure comb formation, the Raman gain must then be suppressed by more than $130\times$. This corresponds to $\delta/2\pi = 342$ GHz for diamond, indicating that an FSR $>$ 684 GHz is required. Similarly, we calculate the gain coefficient at telecommunications wavelengths, taking into account the wavelength dependence of both the Kerr and Raman gain \cite{Latawiec,Savitski,Mildren}. At a pump wavelength of 1560 nm, the gain coefficients are $g_K = 0.1$ cm/GW and $g_R = 6$ cm/GW, which requires an FSR $>$ 464 GHz to suppress Raman gain. The FSR threshold is lower at 1560 nm since the Raman gain coefficient decreases more rapidly with increasing wavelength as compared to the Kerr gain coefficient \cite{Savitski}. 

\begin{figure}[tbp]
\centerline{\includegraphics[width=7.5cm]{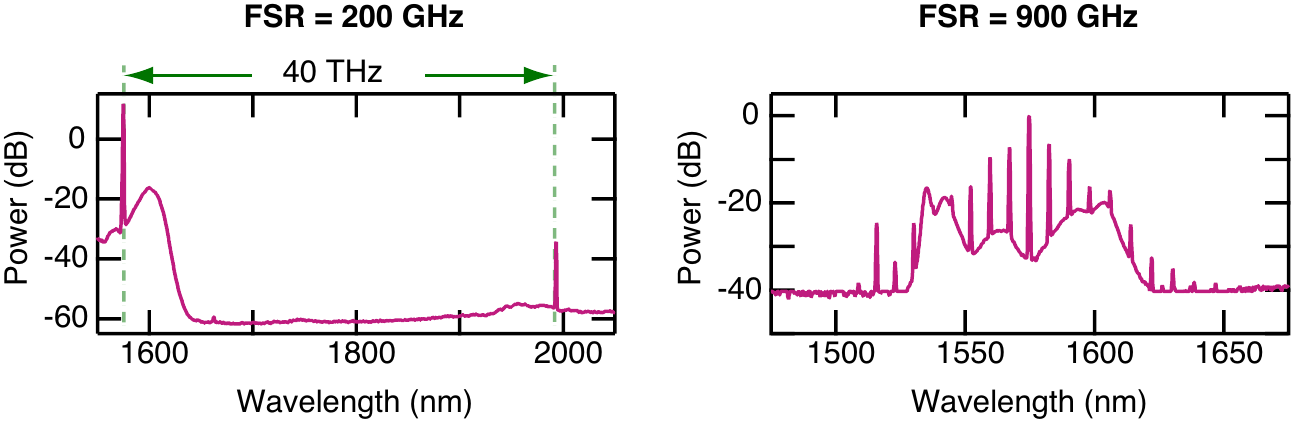}}
\caption{Experimental spectrum for comb generation in diamond microresonators pumping at 1575 nm for two different FSR's of 200 GHz and 900 GHz.}
\label{Fig3}
\end{figure}

We confirm this analysis by performing simulations of the comb generation dynamics using the modified Lugiato-Lefever equation \cite{Lugiato,Haelterman,Matsko,Chembo,Coen,Lamont}, including effects of higher-order dispersion, self-steepening, and the Raman effect. Here, we assume he pump and the Stokes fields are co-polarized. The equation can be written as 


\begin{equation}\label{eq1}
\left.\begin{aligned}
&T_{R}\frac{\partial E(t,\tau)}{\partial t} = \left[-\alpha - i\delta_0 +iL \sum_{n \ge 2}\frac{\beta_n}{n!} \left( i\frac{\partial}{\partial \tau} \right)^n\right.\\
&+ \left. i\gamma L\left(1+\frac{i}{\omega_0}\frac{\partial}{\partial \tau}\right)\int_0^{\infty}R(t')\big|E(t-t',\tau)\big|^2dt'\right]E(t,\tau) \\ 
&+ \sqrt{\kappa} E_{\text{in}},
\end{aligned}\right.
\end{equation}

\noindent 
where $T_R$ is the round-trip time, $E(t,\tau)$ is the field in the microresonator, $\alpha$ is the round-trip loss, $\delta_0$ is the pump-cavity detuning, $\beta_n$ is the $n$th order dispersion coefficient, $\gamma$ is the nonlinear parameter, $R(t)$ is the Raman response, $\kappa$ is the power transmission coefficient of the bus-waveguide-microresonator coupling region, and $E_{\text{in}}$ is the input field. We simulate the dispersion of the waveguide using a finite-element mode solver. The waveguide cross section of the diamond microresonator is 300$\times$400 nm which allows for anomalous GVD at 790 nm. Figure \ref{Fig2}(a) shows the simulated spectra for two different FSR's, 787 GHz and 658 GHz, respectively. The dispersion, propagation loss (0.5 dB/cm), and pump power (300 mW) are the same for the two FSR's. The spacings of the red circles in the spectral plots correspond to the Raman frequency shift. Depending on the FSR, we observe two different final states, corresponding to Kerr comb formation [Fig. \ref{Fig2}(a) top] and Raman oscillation [Fig. \ref{Fig2}(a) top]. Our simulations agree well with the calculated FSR required for pure Kerr comb generation. The larger FSR microresonator allows for suppression of the Raman gain enabling Kerr combs and soliton formation [Fig. \ref{Fig2}(a) inset], whereas the smaller FSR microresonator does not suppress the Raman gain, resulting in Raman oscillation becoming dominant and destabilizing the modelocking process of Kerr frequency combs. We also observe comb lines between the Raman peaks, which we attribute to Raman-assisted FWM. We observe this threshold behavior for Kerr comb formation for other waveguide cross sections provided that the GVD is anomalous and is largely dependent on the microresonator FSR. In addition, we simulate the dynamics for 1560-nm pump using a waveguide cross section of 700$\times$800 nm which allows for anomalous GVD at the pump. Figure \ref{Fig2}(b) shows the simulated spectra for two different FSR's, 500 GHz and 417 GHz, respectively. Our simulations show that, similar to the 790 nm pump case, a Kerr comb is generated for the 500 GHz FSR, but Raman oscillation dominates for 417 GHz, which agrees well with our threshold calculation. Figure \ref{Fig3} shows experimental results showing the two different regimes below and above the FSR threshold (200 GHz and 900 GHz) pumping at 1575 nm. The waveguide cross section is 700$\times$800 nm. Raman oscillation is observed 200-GHz FSR, and Kerr comb formation is observed without Raman effects for 900-GHz FSR \cite{Hausmann,Latawiec}. We conclude that choosing a large FSR is critical for Raman suppression and generation of a pure Kerr comb, particularly at lower wavelengths near visible where the Raman gain is large. The FSR threshold of 684 GHz at 800 nm corresponds to a microring radius of 26 $\mu$m. For such small device dimensions, the contribution from bending losses could increase, leading to lower quality factor devices, and the large FSR can be detrimental to various applications that require smaller comb spacings. 

\begin{figure}[tbp]
\centerline{\includegraphics[width=7.0cm]{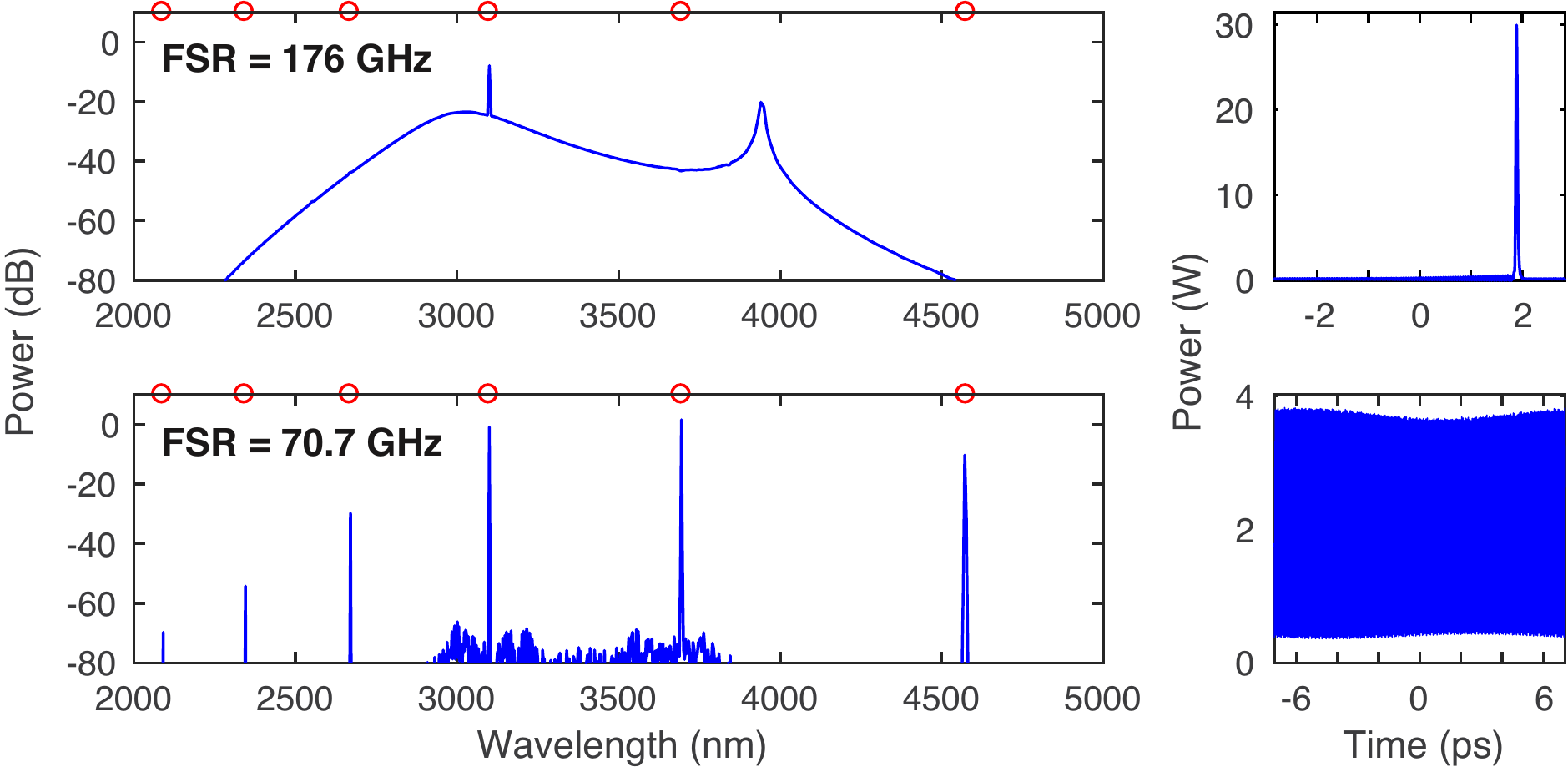}}
\caption{Simulated spectrum (left) and temporal profile (right) for comb generation in silicon microresonators pumping at 3100 nm for FSR's of 176 GHz (top) and 70.7 GHz (bottom).}
\label{Fig4}
\end{figure}

We perform a similar analysis for silicon microresonators operating in the mid-IR. For a 3100 nm pump wavelength, the Kerr and Raman gain coefficients are $g_K = 1.21$ cm/GW and $g_R = 3.05$ cm/GW, respectively. The smaller Raman gain coefficient results in a smaller ratio between the two gain coefficients $g_R/g_K$. Thus, the suppression of Raman for pure Kerr comb formation requires an FSR $>$ 167 GHz. We consider an etchless microresonator with a cross section of 500$\times$1400 nm which allows for anomalous GVD at the pump wavelength of 3100 nm. Similar to our diamond analysis, we fix the dispersion, propagation loss (0.5 dB/cm), and pump power (100 mW) for each FSR we investigate. Figure \ref{Fig4} shows the simulated spectra and corresponding temporal profiles for two different FSR's of 176 GHz and 70.7 GHz. We observe pure Kerr comb formation with minimal Raman effects for the large FSR which again agrees with the threshold calculation. For 70.7 GHz, the Raman effect dominates and we only see Raman oscillation for this pump power. For FSR's between 176 GHz and 70.7 GHz, we observe Raman and Kerr oscillation in the same device for different pump-cavity detunings. We believe this behavior can be explained by the Raman-assisted FWM effect which enables parametric oscillation and Kerr comb formation to occur.

\begin{figure}[tbp]
\centerline{\includegraphics[width=7.0cm]{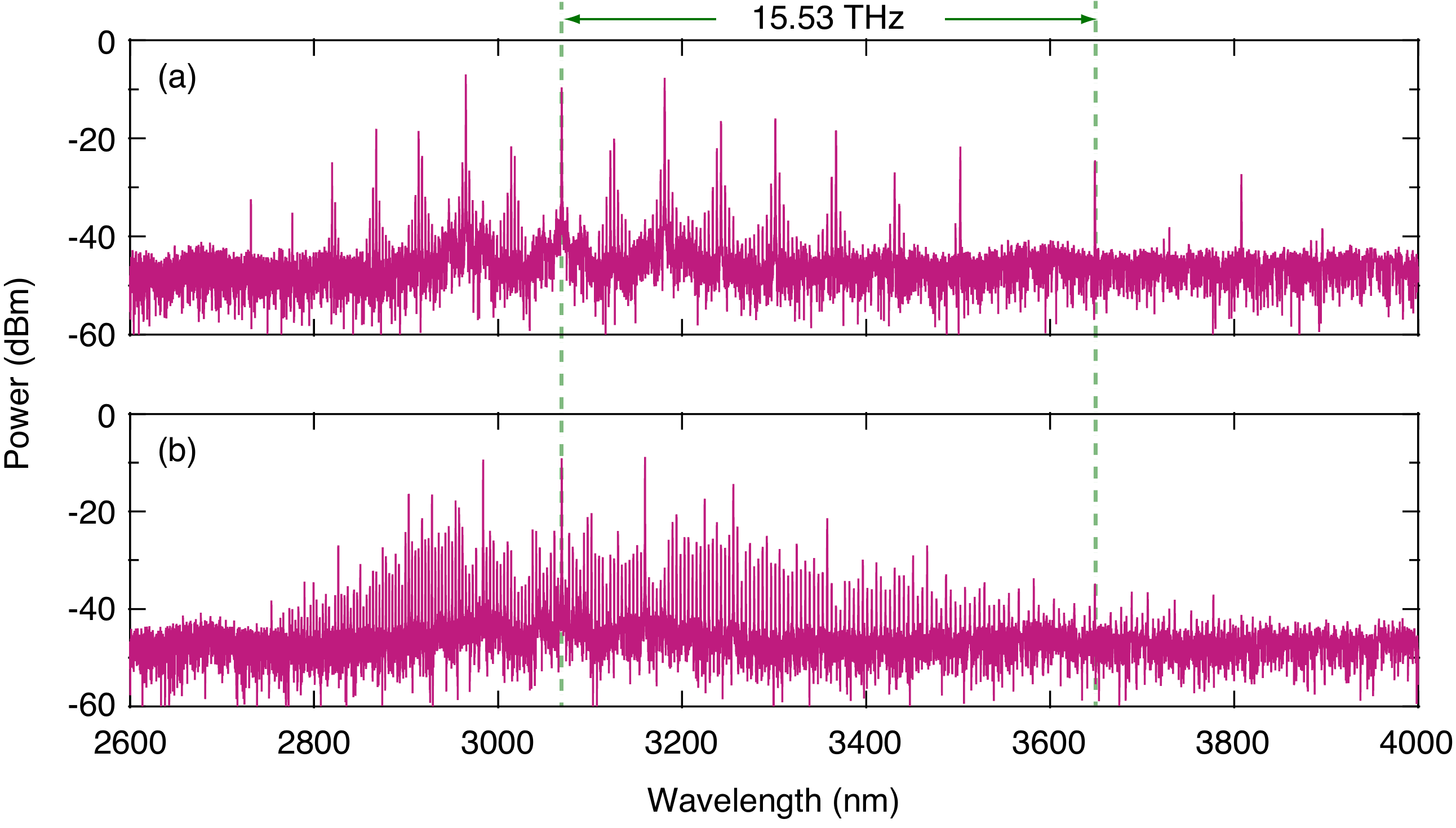}}
\caption{Experimental spectra for comb generation in a silicon microresonator pumping at 3100 nm (a) with Raman oscillation and (b) without Raman effects. The FSR is 128 GHz, which positions the Raman gain peak between the two cavity resonances of the microresonator.} 
\label{Fig5}
\end{figure}

We experimentally investigate comb generation in a silicon microresonator with a waveguide cross section of 500$\times$1400 nm \cite{Yu,Griffith} and an FSR of 128 GHz, which is chosen here to allow for the Raman gain peak to lie between two cavity resonances which minimizes the Raman effect. Figure \ref{Fig4} shows the two different realizations of combs generated in the same microresonator for 80 mW of pump power in the bus waveguide. By choosing a different frequency detuning between the pump and the cavity resonance, we generate a comb state with strong Raman oscillation [Fig. \ref{Fig5} (a)] and a comb state absent of strong Raman effects [Fig. \ref{Fig5}(b)] in the same microresonator. The strong comb lines in Fig. \ref{Fig5}(a) correspond to 1/9\textsuperscript{th} of the Raman frequency shift. In contrast, the strong comb lines in Fig. \ref{Fig5}(b) do not match with the Raman shift and is a result of multiple-soliton formation \cite{Yu}. Our experimental results indicate that in the intermediate regime between the FSR's of 70.7 and 176 GHz there exists competition between Raman and Kerr effects resulting in the existence of two different states.    

\begin{figure}[tbp]
\centerline{\includegraphics[width=7.5cm]{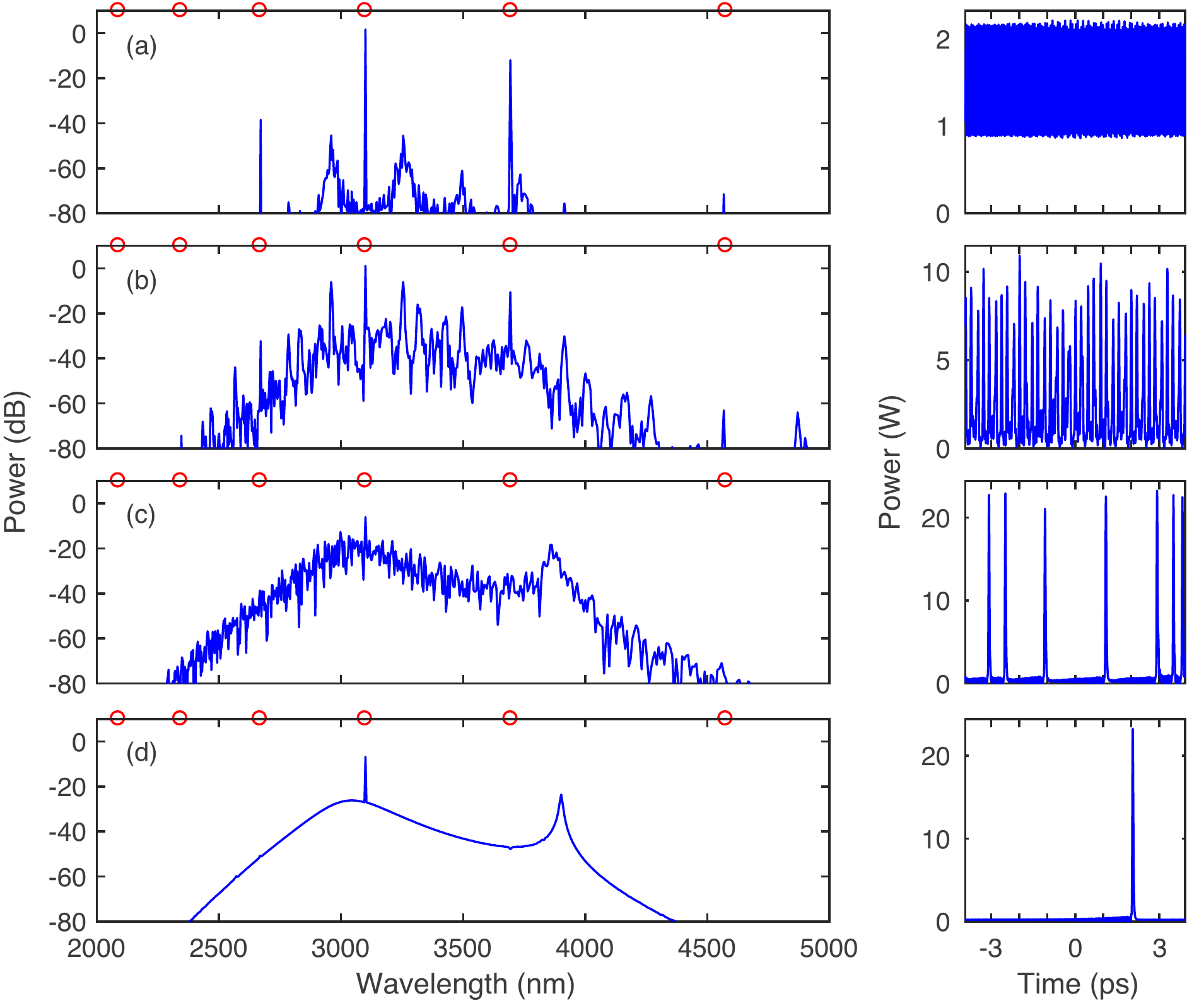}}
\caption{Simulated spectrum (left) and temporal profile (right) for comb generation in silicon microresonators pumping at 3100 nm for FSR of 129 GHz for pump detunings of (a) -0.9 GHz, (b) 0 GHz, (c) 3.5 GHz, and (d) 4.1 GHz with respect to the cold cavity resonance.}
\label{Fig6}
\end{figure}

We numerically investigate this intermediate regime in silicon where competition between Raman and Kerr gain occurs. Figure \ref{Fig6} shows the comb generation dynamics for a 129 GHz FSR silicon microresonator as the pump-cavity detuning is increased. In Fig. \ref{Fig6}(a) we see that, as the pump power builds up, Raman oscillation first occurs. As the detuning is increased, Kerr oscillation occurs as well [Fig. \ref{Fig6}(b)]. Furthermore, we observe non-degenerate FWM driven by the pump and Stokes fields. With further detuning, the Raman lines diminish and multi-soliton formation occurs [Fig. \ref{Fig6}(c)]. Figure \ref{Fig6}(d) shows the final detuning step where single soliton formation occurs and the Raman effect is absent. Thus, with sufficient pump powers, Kerr comb formation is still possible in silicon. In this intermediate regime, the Raman effect diminishes and Kerr oscillation dominates for higher FSR's, and the Raman effect dominates for lower FSR's. This intermediate regime in silicon exists due to the fact that the Raman and Kerr gain are comparable. However, in diamond, this intermediate range is minimal, as the Raman gain is much larger compared to the Kerr gain.

In conclusion, due to the large Raman effect in silicon and diamond, relatively large FSR microresonators are required for mode-locked Kerr comb generation. While silicon offers some flexibility as the Raman and Kerr gain are comparable, the microresonator cavity length must be carefully chosen in diamond due to the high Raman gain, especially at shorter pump wavelengths. Our current investigation considers only one polarization mode for the optical field and does not take into account the polarization/crystal-orientation dependence of the Raman process in crystalline materials. For example, in diamond, for certain waveguide propagation directions the Kerr and Raman effects can be segregated into two orthogonal polarizations \cite{Mildren}. Our results indicate that such approaches utilizing the tensorial nature of the Raman process must be employed in order to realize lower FSR combs that are needed for applications such as astronomical spectroscopy and optical clocks.

\textbf{Funding.} Defense Advanced Research Projects Agency (DARPA) (W31P4Q-15-1-0015); Air Force Office of
Scientific Research (AFOSR) (FA9550-15-1-0303); National Science Foundation (NSF) (ECS-0335765, ECCS-1306035).
\\
\\
\textbf{Acknowledgment.} This work was performed in part at the Cornell Nano-Scale Facility and Center for Nanoscale Systems at Harvard, both members of the National Nanotechnology Infrastructure Network, supported by the NSF. 



\begin{thebibliography}{99}
  \bibitem{Kippenberg}
      T. J. Kippenberg, R. Holzwarth, and S. A. Diddams, Science \textbf{332}, 555 (2011). 

      \bibitem{Del'Haye}
P. Del'Haye, A. Schliesser, O. Arcizet, T. Wilken, R. Holzwarth, and T. J. Kippenberg, Nature, \textbf{450}, 1214 (2007).
  
  \bibitem{Savchenkov} 
  A. A. Savchenkov, A. B. Matsko, V. S. Ilchenko, I. Solomatine, D. Seidel, and L. Maleki, Phys. Rev. Lett. \textbf{101,} 093902 (2008).

        \bibitem{Levy}
J. S. Levy, A. Gondarenko, M. A. Foster, A. C. Turner-Foster, A. L. Gaeta, and M. Lipson, Nat. Photonics, \textbf{4}, 37 (2010).

        \bibitem{Razzari}
L. Razzari, D. Duchesne, M. Ferrera, R. Morandotti, S. Chu, B. E. Little and D. J. Moss, Nat. Photonics \textbf{4}, 41 (2010). 
  
    \bibitem{Ferdous} F. Ferdous, H. Miao, D. E. Leaird, K. Srinivasan, J. Wang, L. Chen, L. T. Varghese, and A. M. Weiner, Nat. Photonics  \textbf{5}, 770 (2011).
  
\bibitem{Papp} S. B. Papp and S. A. Diddams, Phys. Rev. A \textbf{84}, 053833 (2011).

            \bibitem{Jung}
H.~Jung, C.~Xiong, K.~Y. Fong, X.~Zhang, and H.~X. Tang, Opt. Lett. \textbf{38}, 2810 (2013).
  
        \bibitem{Hausmann}
B. J. M. Hausmann, I. Bulu, V. Venkataraman, P. Deotare, and M. Lon\v{c}ar, Nat. Photonics \textbf{8}, 369 (2014).  

\bibitem{Xue}
X. Xue, Y. Xuan, Y. Liu, P.-H. Wang, S. Chen, J. Wang, D. E. Leaird, M. Qi, and A. M. Weiner, Nat. Photonics \textbf{9}, 594 (2015). 

\bibitem{Huang}
S.-W. Huang, H. Zhou, J. Yang, J. F. McMillan, A. Matsko, M. Yu, D.-L. Kwong, L. Maleki, and C. W. Wong, Phys. Rev. Lett. \textbf{114}, 053901 (2015).

\bibitem{Yu}
M. Yu, Y. Okawachi, A. G. Griffith, M. Lipson, and A. L. Gaeta, Optica \textbf{3}, 854 (2016). 

\bibitem{Yi}
X. Yi, Q.-F. Yang, K. Y. Yang, and K. Vahala, Opt. Lett. \textbf{41}, 3419 (2016).  

\bibitem{Karpov}
M. Karpov, H. Guo, A. Kordts, V. Brasch, M. H. P. Pfeiffer, M. Zervas, M. Geiselmann, and T. J. Kippenberg, Phys. Rev. Lett. \textbf{116}, 103902 (2016).

\bibitem{Diddams07}
S. A. Diddams, L. Hollberg, and V. Mbele, Nature \textbf{445}, 627 (2007).

\bibitem{Udem}
T. Udem, R. Holzwarth, and T. W. Hansch, Nature \textbf{416}, 233 (2002).

\bibitem{LevyPTL}
J. S. Levy, K. Saha, Y. Okawachi, M. A. Foster, A. L. Gaeta, and M. Lipson, IEEE Photonics Technol. Lett. 24, 1375 (2012).

\bibitem{Koos}
J. Pfeifle, V. Brasch, M. Lauermann, Y. Yu, D. Wegner, T. Herr, K. Hartinger, P. Schindler, J. Li, D. Hillerkuss, R. Schmogrow, C. Weimann, R. Holzwarth, W. Freude, J. Leuthold, T. J. Kippenberg, and C. Koos, Nat. Photonics 8, 375 (2014).

\bibitem{Saha}
K. Saha, Y. Okawachi, B. Shim, J. S. Levy, R. Salem, A. R. Johnson, M. A. Foster, M. R. E. Lamont, M. Lipson, and A. L. Gaeta, Opt. Express \textbf{21}, 1335 (2013).

\bibitem{Herr}
T. Herr, V. Brasch, J. D. Jost, C. Y. Wang, N. M. Kondratiev, M. L. Gorodetsky, and T. J. Kippenberg, Nat. Photonics \textbf{8}, 145 (2014).

\bibitem{Yi15}
X. Yi, Q.-F. Yang, K. Y. Yang, M.-G. Suh, and K. Vahala, Optica \textbf{2}, 1078 (2015).

\bibitem{Joshi}
C. Joshi, J. K. Jang, K. Luke, X. Ji, S. A. Miller, A. Klenner, Y. Okawachi, M. Lipson, and A. L. Gaeta, Opt. Lett. \textbf{41}, 2565 (2016).

\bibitem{Min}
B. Min, L. Yang, and K. J. Vahala, Appl. Phys. Lett. \textbf{87}, 181109 (2005). 

\bibitem{Agha}
I. H. Agha, Y. Okawachi M. A. Foster, J. E. Sharping, and A. L. Gaeta, "Four-wave-mixing parametric oscillations in dispersion-compensated high-Q silica microspheres," Phys. Rev. A \textbf{76}, 043837 (2007).

\bibitem{Grudinin}
I. S. Grudinin, L. Baumgartel, and N. Yu, Opt. Express \textbf{21}, 26929 (2013).

\bibitem{Liang}
W. Liang,  A. B. Matsko,  A. A. Savchenkov, V. S. Ilchenko, D. Seidel, and L. Maleki, In Frequency Control and the European Frequency and Time Forum (FCS), 2011 Joint Conference of the IEEE International, pp 1-6, May 2011.

\bibitem{Griffith}
A. G. Griffith, M. Yu, Y. Okawachi, J. Cardenas, A. Mohanty, A. L. Gaeta, and M. Lipson, Opt. Express \textbf{24}, 13044 (2016). 

\bibitem{Latawiec}
P. Latawiec, V. Venkataraman, M. J. Burek, B. J. M. Hausmann, I. Bulu, and M. Lon\v{c}ar, Optica \textbf{2}, 924 (2015). 

\bibitem{Kumar}
S. Kumar and S. K. Biswas, J. Opt. Soc. Am. B \textbf{33}, 1677 (2016).

\bibitem{Liu}
X. Liu, C. Sun, B. Xiong, L. Wang, J. Wang, Y. Han, Z. Hao, H. Li, Y. Luo, J. Yan, T. Wei, Y. Zhang, and J. Wang, arXiv:1611.01994.

\bibitem{Yang}
Q.-F. Yang, X. Yi, K. Y. Yang, and K. Vahala, Nat. Phys. \textbf{13}, 53 (2017).

\bibitem{Savitski}
V. G. Savitski, S. Reilly, and A. J. Kemp, IEEE J. Quant. Electron. \textbf{49}, 218 (2013). 

\bibitem{Yu_arXiv}
M. Yu, Y. Okawachi, A. G. Griffith, N. Picqu\'e, M. Lipson, and A. L. Gaeta, arXiv:1610.01121.

\bibitem{Gai}
X. Gai, Y. Yu, B. Kuyken, P. Ma, S. J. Madden, J. Van Campenhout, P. Verheyen, G. Roelkens, R. Baets, and B. Luther-Davies Laser Photon. Rev. \textbf{7}, 1054 (2013). 

\bibitem{Raghunathan}
V. Raghunathan, D. Borlaug, R. R. Rice, and B. Jalali, Opt. Express \textbf{15}, 14355 (2007). 

\bibitem{Pask}
H. M. Pask, Prog. in Quantum Electron. \textbf{27}, 3 (2003). 

\bibitem{Mildren}
R. P. Mildren, \emph{Optical Engineering of Diamond} (Wiley-VCH, 2013). 

\bibitem{Lugiato} 
L. A. Lugiato and R. Lefever, Phys. Rev. Lett. \textbf{58}, 2209 (1987).

\bibitem{Haelterman} 
M. Haelterman, S. Trillo, and S. Wabnitz, Opt. Comm. \textbf{91}, 401 (1992).

\bibitem{Matsko} 
A. B. Matsko, A. A. Savchenkov, W. Liang, V. S. IlChenko, D. Seidel, and L. Maleki, Opt. Lett. \textbf{36}, 2845 (2011).

\bibitem{Chembo} 
Y. K. Chembo and C. R. Menyuk, Phys. Rev. A \textbf{87}, 053852 (2013).

\bibitem{Coen} 
S. Coen and M. Erkintalo, Opt. Lett. \textbf{38}, 1790 (2013).

\bibitem{Lamont}
M. R. E. Lamont, Y. Okawachi, and A. L. Gaeta, Opt. Lett. \textbf{38}, 3478 (2013). 
\end{thebibliography}
\end{document}